\documentclass{dfsnet}

\dfsnetcameraready

\usepackage{amsmath,graphicx}
\usepackage{latexsym}
\usepackage{amsfonts,amssymb}
\usepackage{stfloats}
\usepackage{url}
\usepackage{lipsum}
\usepackage{upgreek}
\usepackage{hhline}
\usepackage{enumerate}
\usepackage{mathtools}
\usepackage{booktabs}
\usepackage{graphicx}

\usepackage{multirow}
\usepackage{makecell}
\usepackage{colortbl}
\usepackage{booktabs,siunitx}
\usepackage{tabularx}
\usepackage[none]{hyphenat}

\newcommand{\cmark}{\checkmark}%
\newcommand{\xmark}{\small$\boldsymbol\times$}%
\DeclarePairedDelimiter\floor{\lfloor}{\rfloor}

\title{DFSNet: A Steerable Neural Beamformer Invariant to Microphone Array Configuration for Real-Time, Low-Latency Speech Enhancement}
\name{Anton Kovalyov, Kashyap Patel, Issa Panahi}
\address{Electrical and Computer Engineering, University of Texas at Dallas, Richardson, TX, USA}
\email{anton.kovalyov@utdallas.edu}

\begin{document}

\maketitle
 
\begin{abstract}
Invariance to microphone array configuration is a rare attribute in neural beamformers. Filter-and-sum (FS) methods in this class define the target signal with respect to a reference channel. However, this not only complicates formulation in reverberant conditions but also the network, which must have a mechanism to infer what the reference channel is. To address these issues, this study presents Delay Filter-and-Sum Network (DFSNet), a steerable neural beamformer invariant to microphone number and array geometry for causal speech enhancement. In DFSNet, acquired signals are first steered toward the speech source direction prior to the FS operation, which simplifies the task into the estimation of delay-and-summed reverberant clean speech. The proposed model is designed to incur low latency, distortion, and memory and computational burden, giving rise to high potential in hearing aid applications. Simulation results reveal comparable performance to noncausal state-of-the-art.\par
\end{abstract}

\section{Introduction}
With recent advancements in deep learning, deep neural network (DNN)-based beamformers, also known as neural beamformers, have gained considerable traction in the literature \cite{adl_mvdr,beam_tasnet,sep_interaction,css}. Neural beamformers are known to outperform both statistical and DNN-based single-channel methods on different tasks. Time-domain methods \cite{temporal_spatial,dsenet,uxnet} are an increasingly popular class among neural beamformers because of their high potential in latency-demanding applications, such as hearing aids. However, unless retrained, the proposed networks rarely provide invariance to microphone array configuration, i.e., microphone number and array geometry, an attribute of special importance in ad-hoc array scenarios.\par

The Filter-and-Sum Network (FaSNet) systems \cite{fasnet, fasnet_tac} are state-of-the-art (SOTA) in time-domain neural beamformers suitable for ad-hoc arrays. FaSNet is an end-to-end system that performs framewise filter-and-sum (FS) beamforming in the time domain. Consistent with other multi-channel methods, FaSNet specifies its target signal with respect to a reference microphone. Thus, when trained for speech enhancement (SE), the target signal of FaSNet is the clean reverberant speech at a reference microphone. However, this formulation introduces two complications. (1) The model needs to somehow learn how to combine the different-channel signals to both reduce noise as well as reconstruct the direct path and reverberant components of speech at a reference microphone. (2) As a consequence of array geometry invariance, special processing with respect to the reference microphone must be introduced, otherwise the model has no means to infer what the reference microphone is.\par

Motivated by the above observations, this study proposes Delay-Filter-and-Sum Network (DFSNet), a steerable neural beamformer invariant to microphone array configuration for real-time, low-latency SE. DFSNet operates in a framewise manner and follows a linear signal model analogous to frequency-domain FS beamforming. In the proposed model, time-domain waveforms are first delayed by a set of integer and fractional delay finite impulse response (FIR) filters toward the speech source direction. Delayed signals are then converted into a latent space representation through a linear transformation. Next, masks for each channel are estimated by a stack of recurrent channel interaction (RCI) blocks, which efficiently combine recurrent processing with a channel interaction (CI) technique similar to transform-average-concatenate (TAC) \cite{fasnet_tac}. Finally, FS is applied in the latent space representation followed by a linear transformation to convert the result back to the time domain. As a consequence of signal delay prior to FS, the target signal of DFSNet is defined as the delay-and-sum (DS) clean reverberant speech. With this approach, DFSNet simplifies the task into learning how to collectively reduce noise at individual channels; avoids specifying a reference microphone; and allows steering to different directions without retraining.\par 

DFSNet is benchmarked against SOTA, including causal and noncausal FaSNet variants. Results show that the proposed method approaches and sometimes exceeds the performance of noncausal systems. An ablation study is also conducted.\par

\section{Problem Formulation}
Let us consider an array of $C$ microphones and arbitrary geometry in a reverberant environment. The time-domain signal captured by the $c$-th microphone is modeled by
\begin{equation} \label{eq:noisy}
    \mathbf{y}_c = \mathbf{x}_c + \mathbf{v}_c \, , \quad c = 1, 2, \ldots, C\,,
\end{equation}
where $\mathbf{x}_c$ denotes clean reverberant speech and $\mathbf{v}_c$ is noise. Let $\mathbf{u}$ and $\mathbf{m}_c$ be the 3-dimensional (3D) positions of the speech source and $c$-th microphone, respectively. The time difference of arrival (TDOA) in samples of the signal originating at $\mathbf{u}$ when received between $\mathbf{m}_1$ and $\mathbf{m}_i$, for $i = 2,3,\ldots,C$, is given by
\begin{equation} \label{eq:tdoa}
    \tau_{i} = f s^{-1} \left( ||\mathbf{u} - \mathbf{m}_1|| - ||\mathbf{u} - \mathbf{m}_i|| \right),
\end{equation}
where $f$ is the sampling rate and $s$ is the propagation speed. We set $\mathbf{m}_1$ to the furthest microphone position from source. Let $\tilde{\mathbf{\tau}}_i$ be a known positive estimate of $\tau_{i}$. We can align the acquired signals toward an approximate direction of the speech source by
\begin{equation} \label{eq:align}
    \mathbf{y}_i^a = \left( \mathbf{h}_{F_i} \ast \mathbf{h}_{D_i} \right) \ast \mathbf{y}_i \,,\quad i = 2,3,\ldots,C,
\end{equation}
where $\mathbf{h}_{D_i}$ and $\mathbf{h}_{F_i}$ are causal integer and fractional delay FIR filters, respectively, and $\ast$ denotes convolution. The subscripts $D_i = \floor{\tilde{\tau}_i}$ and $F_i = \tilde{\tau_i} - D_i$ specify the sample delay of a filter $\mathbf{h}$. Implementation of $\mathbf{h}_{D_i}$ is trivial, whereas for $\mathbf{h}_{F_i}$, we employ sinc-based fractional delay FIR filters \cite{frac_delay} of equal length $M$. The latter incur a fixed integer latency $D = \lfloor\frac{M-1}{2}\rfloor$. Hence, $\mathbf{y}_1$ is also delayed\footnote{This delay can be reduced in a variable manner by adjusting $\mathbf{h}_{D_i}$ to also reflect upon latency incurred by fractional delay filtering.} to compensate for this latency by
\begin{equation} \label{eq:align_ref}
    \mathbf{y}_1^a = \mathbf{h}_D \ast \mathbf{y}_1 \,,
\end{equation}
Next, let us consider the causal DS beamformer in
\begin{equation} \label{eq:ds}
    \mathbf{y}_{DS} = \dfrac{1}{C}\sum_{c=1}^{C} \mathbf{y}_c^a \,.
\end{equation}
Applying (\ref{eq:noisy}), (\ref{eq:align}) and (\ref{eq:align_ref}), we note that $\mathbf{y}_{DS}$ can be separated into its speech component 
\begin{equation} \label{eq:target}
    \mathbf{x}_{DS} = \frac{1}{C} \left[ \mathbf{h}_D \ast \mathbf{x}_1 + \sum_{i=2}^C \left( \mathbf{h}_{F_i} \ast \mathbf{h}_{D_i} \right) \ast \mathbf{x}_i \right]
\end{equation}
and similarly defined noise component $\mathbf{v}_{DS}$. The problem is formulated as causal estimation of $\mathbf{x}_{DS}$.\par

\section{Delay-Filter-and-Sum Network (DFSNet)}
\begin{figure*}[t!]
    \centering
    \includegraphics[trim={0 0 0 0},clip, width=1\textwidth]{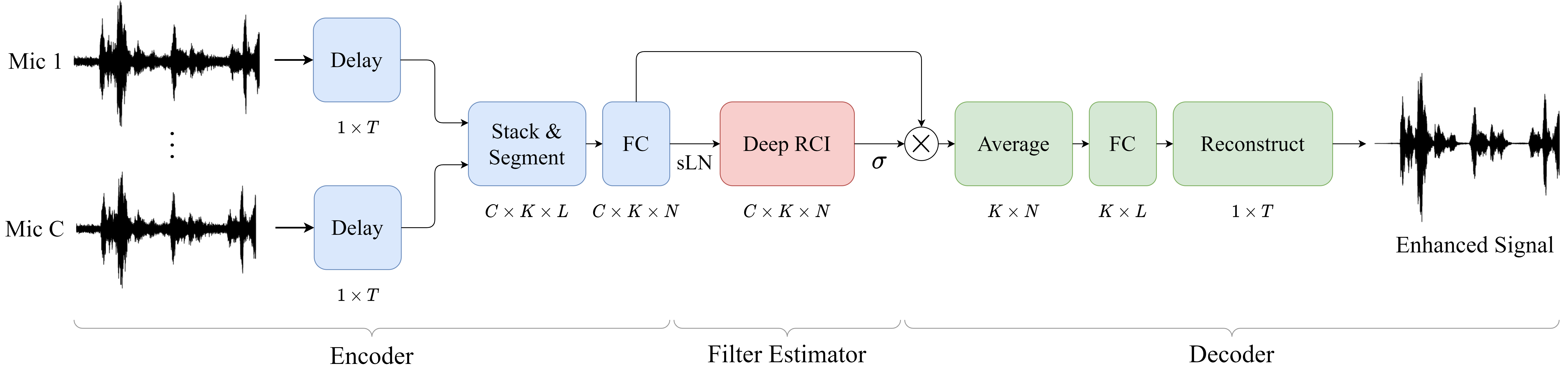}
    \caption{System flowchart of the proposed DFSNet.}
    \label{fig:model}
\end{figure*}

As shown in Fig. \ref{fig:model}, the processing pipeline of DFSNet consists of three stages: encoder, filter estimator, and decoder.\par

\subsection{Encoder}
At the encoder, input channels are first aligned applying (\ref{eq:align}) and (\ref{eq:align_ref}) to produce utterances $\mathbf{y}^a_c$ of length $T$ samples. Next, each utterance is segmented into $K$ sequential overlapping frames of length $L$ samples and 50\% overlap. Let $\mathbf{y}^a_{c,k} \in \mathbb{R}^{1 \times L}$ be a segment corresponding to channel $c$ and frame index $k$, for $k = 1,2,\ldots,K$. A linear transformation is then applied to convert each $\mathbf{y}^a_{c,k}$ into an $N$-dimensional latent space representation
\begin{equation} \label{eq:encoder}
    \mathbf{z}_{c,k} = \mathbf{y}^a_{c,k} \mathbf{B}_e \,,
\end{equation}
where $\mathbf{B}_e \in \mathbb{R}^{L \times N}$ are weights of a fully connected (FC) layer.\par

\subsection{Filter Estimator}
The filter estimator estimates channel and time-varying filters given by mask vectors $\hat{\mathbf{m}}_{c,k} \in \mathbb{R}^{1 \times N}$ for application in the latent space corresponding to (\ref{eq:encoder}). In this module, each $\mathbf{z}_{c,k}$ is first normalized applying sliding window layer normalization (sLN) to reduce variability and speed up training, followed by stacked RCI blocks and a sigmoid nonlinearity to ensure nonnegative masks. Both sLN and RCI are proposed here and described separately in Sections \ref{sec:sln} and \ref{sec:rci}, respectively.\par

\subsection{Decoder}
At the decoder, estimated masks $\hat{\mathbf{m}}_{c,k}$ and latent space representations $\mathbf{z}_{c,k}$ are multiplied and summed across the channel dimension followed by transformation back to the time domain by an FC layer with weights $\mathbf{B}_d \in \mathbb{R}^{N \times L}$ and no bias. The complete procedure is given by
\begin{equation} \label{eq:decoder}
    \hat{\mathbf{x}}_{DS,k} = \left( \dfrac{1}{C}\sum_{c=1}^C  \hat{\mathbf{m}}_{c,k} \odot \mathbf{z}_{c,k} \right) \mathbf{B}_d \,,
\end{equation}
where $\odot$ denotes element-wise product. An estimate of $\mathbf{x}_{DS}$ is then reconstructed by the overlap-add operation.\par

The proposed encoder/decoder operations follow a linear signal model analogous to frequency-domain FS beamforming, with the difference that instead of short-time Fourier transform (STFT), we apply forward and inverse transformations learned by the network. A linear signal model is preferred here since it is not as likely to cause unpleasant distortions as its nonlinear counterpart. Moreover, this model can be paired with distortion control schemes \cite{distortion_control} to behave similarly to a minimum variance distortionless response (MVDR) beamformer \cite{FD-MVDR}, thus making it especially suitable for hearing aid applications.\par 

\subsection{Sliding Window Layer Normalization (sLN)} \label{sec:sln}
The proposed sLN is similar to cumulative layer normalization (cLN) \cite{convtasnet} with the difference that normalization is performed over a sliding window of fixed size $R$ rather than cumulatively, thus allowing for better adaptation in applications where signal statistics can drastically change over time. The proposed sLN is applied at each channel independently as follows
\begin{equation} \label{eq:sln}
    \begin{aligned}
        \text{sLN}\left(\mathbf{f}_{c,k}\right) &= \dfrac{\mathbf{f}_{c,k} - \mu_{c,k}\mathbf{1}}{\sqrt{\sigma_{c,k}^2 + \epsilon}} \odot \boldsymbol{\gamma} + \boldsymbol{\beta} \\
        \mu_{c,k} &= \dfrac{1}{N R_{k}} \sum_{t = 1}^{R_{k}} \sum_{n=1}^N f_{c,k-t+1,n} \\
        \sigma^2_{c,k} &= \dfrac{1}{N R_{k}} \sum_{t = 1}^{R_{k}} \sum_{n=1}^N \left( f_{c,k-t+1,n} - \mu_{c,k} \right)^2 \\
        R_{k} &= \min{ \left\{ k, R \right\} } \,,
    \end{aligned}
\end{equation}
where, $\mathbf{f}_{c,k} \in \mathbb{R}^{1 \times N}$ is a channel and time dependent input vector, $\mathbf{1} \in \mathbb{R}^{1 \times N}$ is a vector of ones, $\boldsymbol{\gamma} \in \mathbb{R}^{1 \times N}$ and $\boldsymbol{\beta} \in \mathbb{R}^{1 \times N}$ are learnable parameters, $\mu_{c,k}$ and $\sigma^2_{c,k}$ are sliding mean and variance computed across time and feature dimensions, $f_{c,*,n}$ is the $n$-th feature of $\mathbf{f}_{c,*}$, and $R_{k}$ is the window size at time index $k$, which converges to $R$ once $k \geq R$. It follows that for $R = 1$, sLN behaves exactly like layer normalization (LN) \cite{layer_norm}, whereas for $R \geq K$, it becomes cLN. The computational overhead of sLN is negligeble if implemented applying dynamic programming by means of two circular buffers of length $R$ each, maintained at each channel independently. Thus, we only need to select an $R$ that provides a good trade-off between performance and memory cost.\par

\subsection{Recurrent Channel Interaction (RCI)} \label{sec:rci}
The proposed RCI block combines gated recurrent units (GRUs) and a CI technique similar to that in TAC \cite{fasnet_tac} blocks. The aim is to gain spatio-temporal context awareness, necessary for estimation of beamforming filters, without sacrificing invariance to microphone number and array geometry. In an RCI block, channel and time dependent input features $\mathbf{f}^{\text{in}}_{c,k} \in \mathbb{R}^{1 \times N}$ first go through a parametric rectified linear unit activation (PReLU) function \cite{prelu}, resulting in $\mathbf{f}_{c,k}$, followed by averaging across the channel dimension by
\begin{equation} \label{eq:average}
    \bar{\mathbf{f}}_{k} = \dfrac{1}{C}\sum_{c=1}^C \mathbf{f}_{c,k} \,.
\end{equation}
Then, the following sequence of operations is performed independently at every channel. First, $\mathbf{f}_{c,k}$ and $\bar{\mathbf{f}}_{k}$ are uniformly partitioned into $P$ nonoverlapping feature bands, denoted respectively as, $\mathbf{f}_{c,k,p} \in \mathbb{R}^{1 \times N/P}$ and $\bar{\mathbf{f}}_{k,p} \in \mathbb{R}^{1 \times N/P}$, for $p = 1,2,\ldots,P$. Then, for every $p$-th partition, $\mathbf{f}_{c,k,p}$ and $\bar{\mathbf{f}}_{k,p}$ are concatenated and fed to a corresponding GRU layer of $H/P$ units in a parallel manner as follows
\begin{equation} \label{eq:gru}
    \mathbf{f}'_{c,k,p} = \text{GRU}_p\left( \left[ \mathbf{f}_{c,k,p}, \bar{\mathbf{f}}_{k,p} \right] \right), \quad p = 1,2,\ldots,P\,,
\end{equation}
where the indexing $p$ in $\text{GRU}_p\left(\cdot\right)$ is used to clarify that we do not include parameter sharing (PS) between GRUs applied at different partitions. The resulting outputs are then concatenated back to form $\mathbf{f}'_{c,k} \in \mathbb{R}^{1 \times H}$, followed by applying an FC layer, with weights $\mathbf{B} \in \mathbb{R}^{H \times N}$ and bias vector $\mathbf{b} \in \mathbb{R}^{1 \times N}$, in sequence with sLN. Finally, we add a skip connection between the output and $\mathbf{f}_{c,k}$ to ease learning in a similar manner as in a ResNet \cite{resnet}. The entire procedure is given by
\begin{equation}
    \mathbf{f}^{\text{out}}_{c,k} = \text{sLN} \left( \mathbf{f}'_{c,k} \mathbf{B} + \mathbf{b} \right) + \mathbf{f}_{c,k} \,.
\end{equation}
The FC layer is used for transformation back to the encoding dimension while allowing communication across the different feature partitions. The purpose of feature partitioning combined with parallel application of $P$ GRU layers with no PS is to evenly reduce both the number of parameters and operations by a factor of $P$. Inspired by the concept of \textit{group convolution} \cite{shufflenet}, we refer to the operation in ({\ref{eq:gru}}) as \textit{group GRU}.\par

\subsection{Local and Global Processing}
Operations in the proposed DFSNet can be divided into \textit{local}, i.e., intra-channel operations such as (\ref{eq:encoder}), (\ref{eq:sln}) and (\ref{eq:gru}); and \textit{global}, i.e., inter-channel operations such as (\ref{eq:average}) and the matrix multiplication in (\ref{eq:decoder}). The parameters involving local operations are shared across channels, whereas states, e.g., hidden states in GRUs, are channel dependent. The lack of reference channel processing is attributed to the channel-alignment procedures in (\ref{eq:align}) and (\ref{eq:align_ref}) combined with the DS target signal definition in (\ref{eq:target}).\par

\subsection{Optimization}
For improved scalability to increasing number of microphones, we want to decrease the ratio of local to global operations. For this purpose, the group GRU operation in (\ref{eq:gru}) can be optimized to compute the GRU's matrix multiplication involving $\bar{\mathbf{f}}_{k,p}$ only once and reuse the result.\par

\section{Experiments}

We evaluate the performance of the proposed DFSNet on the task of SE in a reverberant environment.\par

\subsection{Dataset}
A dataset is generated using clean speech utterances from LibriSpeech \cite{libri} mixed with noise utterances from WHAM! \cite{wham} to simulate noisy speech captured by a microphone array of arbitrary number of microphones and geometry in a reverberant room. The dataset generates 40960, 5120, and 6144, 4-second-long utterances for training, validation, and testing, respectively. The sampling frequency is set to 16 kHz. The training and validation sets are evenly split to consider arbitrary array configurations of 2, 3, 4, 5, and 6 microphones, whereas the test set is evenly split to only consider arbitrary array configurations of 2, 4, and 6 microphones. For each utterance, the dimensions of the room are uniformly sampled between 5 and 10 meters in length and width, and 2 to 4 meters in height. The reverberation time ranges randomly between 0.1 and 0.5 seconds and the sound propagation speed $s$ is fixed to 343 m/s. The 3D microphone positions are randomly selected within 15 cm from the middle of the room. A single speech source along with a randomly varying number between 1 and 4 noise sources are considered. The overall signal-to-noise ratio (SNR) is set to range uniformly between -5 and 15 dB. The different sources are randomly distributed around the room with the constraint of being at least 50 cm away from the walls, and the image method \cite{rir} is applied to compute the corresponding room impulse responses (RIRs). Finally, with the aim of simulating noise in the channel alignment procedure in (\ref{eq:align}) and (\ref{eq:align_ref}) that forms a beam toward the desired source, i.e., the speech source in this particular case, the TDOAs in (\ref{eq:tdoa}) are corrupted to reflect a uniformly and disjointly sampled error between 0 and 5 degrees in azimuth and elevation angles with respect to the source's true position.\par

\subsection{Training and Network Configuration}
DFSNet is trained for 50 epochs with Adam \cite{adam} optimizer and a batch size of 8. The initial learning rate is set to 1e-3 and an exponential decay of 0.98 is applied every epoch. The training objective is given by maximization of scale invariant signal-to-distortion ratio (SI-SDR) \cite{sdr}. The target signal is the delay-and-summed reverberant clean speech $\mathbf{x}_{DS}$ as defined in (\ref{eq:ds}). The frame length $L$ is set to 64 samples, thus causing a latency of 4 ms without counting processing time, which cannot exceed 2 ms. The length of the fractional delay FIR filters $\mathbf{h}_{F_i}$ in (\ref{eq:align}) is set to 17. These filters incur an additional latency of 0.5 ms. The encoding and hidden dimensions $N$ and $H$ are set to 128 and 256, respectively. The number of RCI blocks in the filter estimation module and the number of partitions $P$ in (\ref{eq:gru}) are both set to 4. Finally, the window length $R$ in sLN is set to 1000, which is equivalent to a receptive field of 2 seconds.\par

\subsection{Performance Metrics}
The performance metrics used are: SI-SDR (dB), Perceptual Evaluation of Speech Quality (PESQ) \cite{pesq}, and Short-Time Objective Intelligibility (STOI) \cite{stoi}.\par

\section{Results and Analysis}

\subsection{Comparison with Causal and Noncausal SOTA}
For benchmarking purposes, DFSNet is compared to causal and non-causal SOTA in time-domain models, namely, the causal two-stage FaSNet \cite{fasnet} (FaSNet) and the non-causal single-stage FaSNet with TAC \cite{fasnet_tac} (FaSNet-TAC) multi-channel models, as well as the Convolutional Time Audio Separation Network \cite{convtasnet} (Conv-TasNet) and dual-path recurrent neural network TasNet \cite{dprnn} (DPRNN-TasNet) single-channel models. These are trained under the same conditions as DFSNet. The target signal is the reverberant clean speech at a reference microphone, selected as the closest microphone to source due to its highest SNR. For FaSNet-TAC, Conv-TasNet, and DPRNN-TasNet, we employ, respectively, the best performing configuration in \cite{fasnet_tac}, the causal configuration in \cite{convtasnet}, and the 2-ms-frame-size configuration in \cite{dprnn}. For FaSNet, the same causal configuration in \cite{fasnet} is employed, with the difference that, to compensate for the use of a higher sampling rate, we increase the number of input channels in each convolutional block and the embedding dimension from 64 to 80. For further reference, the well-known frequency-domain MVDR \cite{FD-MVDR} beamformer is also evaluated. For MVDR, we consider the formulation without dereverberation and employ Hann windowing with 50\% overlap, and a frame size of 32 ms. The second order statistics of speech and noise, required by MVDR, are estimated with the actual speech and noise utterances prior mixing.\par

\begin{table*}[t]
    \small
    \centering
    \caption{Comparison with SOTA. Performance measures are computed with respect to clean reverberant speech at closest microphone to source. Bottom of row in DFSNet includes performance results with respect to DS clean reverberant speech.}
    \setlength\tabcolsep{5.1pt}
    \begin{tabular}{c c c c c c c c c}
         \Xhline{2\arrayrulewidth}
         \vspace{-0.27cm} & & & & & & & \\
          Method & \makecell{Multi- \\ channel} & Causal & Latency & \makecell{Model \\ size} & \makecell{GMAC/s \\ (2/4/6 mics)} & \makecell{SI-SDR \\ (2/4/6 mics)} & \makecell{PESQ \\ (2/4/6 mics)} & \makecell{STOI \\ (2/4/6 mics)} \\[0.2cm]
         \Xhline{1\arrayrulewidth} 
         Unprocessed & \xmark & \cmark & 0.1 ms & -- & -- & 5.04/5.04/5.04 & 1.78/1.78/1.78 & 0.75/0.75/0.75 \\
         MVDR & \cmark & \xmark & -- & -- & -- & 7.24/9.13/9.43 & 2.15/2.66/2.91 & 0.83/0.89/0.91 \\
         Conv-TasNet & \xmark & \cmark & 2.0 ms & 5.00M & 5.23 & 10.87/10.89/10.91 & 2.35/2.35/2.35 & 0.84/0.84/0.85 \\
         DPRNN-TasNet & \xmark & \xmark & -- & 2.60M & 5.80 & 12.21/12.37/12.30 & 2.66/2.68/2.66 & 0.87/0.87/0.87 \\
         FaSNet & \cmark & \cmark & 4.0 ms & 1.66M & 1.64/3.29/4.93 & 10.71/11.36/11.45 & 2.24/2.34/2.36 & 0.84/0.86/0.86 \\
         FaSNet-TAC & \cmark & \xmark & -- & 2.76M & 5.29/9.92/14.56 & \textbf{12.87}/\textbf{13.91}/14.22 & \textbf{2.77}/\textbf{2.95}/\textbf{2.99} & \textbf{0.88}/0.90/0.91 \\
         \Xhline{1\arrayrulewidth} 
         DFSNet & \cmark & \cmark & 4.5 ms & \textbf{0.55M} & \textbf{0.50/0.94/1.38} & 
         \makecell{9.38/9.01/8.62 \\ 12.29/13.87/\textbf{14.29}} & \makecell{2.56/2.77/2.82 \\ 2.61/2.88/2.97} & \makecell{0.86/0.89/0.89 \\ 0.87/\textbf{0.91}/\textbf{0.92}} \\
         \Xhline{2\arrayrulewidth}
    \end{tabular}
    \label{tab:comparison}
\end{table*}

Table \ref{tab:comparison} reports the results. GMAC/s specifies a model's Giga Multiply-Accumulate operations per second. We notice that DFSNet underperforms in SI-SDR when evaluated with respect to a reference microphone, especially as the microphone number increases. However, when it comes to perception and intelligibility, DFSNet outperforms all causal methods by a significant margin. Additionally, when evaluated with respect to its target signal, DFSNet outperforms MVDR in all cases, and, with just two microphones, attains comparable performance to the noncausal DPRNN-TasNet. Moreover, when the number of microphones increases, DFSNet approaches and in certain cases exceeds the performance of the noncausal FaSNet-TAC. We further note that DFSNet incurs only a fraction of memory and computational cost of SOTA models, which is largely attributed to the proposed feature partitioning scheme in RCI blocks.\par

\subsection{Ablation study}

\begin{table}[t]
    \small
    \centering
    \caption{Ablation study. When CI is not set, only local features are processed. Unspecified $R$ implies no normalization.}
    \setlength\tabcolsep{3.8pt}
    \begin{tabular}{c c c c c c c}
         \Xhline{2\arrayrulewidth}
         \vspace{-0.3cm} & & & & & & \\
          $P$ & PS & $R$ & CI & \makecell{Model \\ size} & \makecell{GMAC/s \\ (local/global)} & \makecell{SI-SDR \\ /PESQ/STOI} \\[0.2cm]
         \Xhline{1\arrayrulewidth} 
         \vspace{-0.3cm} & & & & & & \\
         1 & -- & 1000 & \cmark & 1.73M & 0.66/0.20 & 13.46/2.80/\textbf{0.90} \\
         4 & \cmark & 1000 & \cmark & 0.25M & 0.22/0.05 & 12.99/2.71/0.89 \\
         4 & \xmark & 1000 & \cmark & 0.55M & 0.22/0.05 & \textbf{13.48}/\textbf{2.82}/\textbf{0.90} \\
         4 & \xmark & 2000 & \cmark & 0.55M & 0.22/0.05 & 13.44/2.79/\textbf{0.90} \\
         4 & \xmark & 1 & \cmark & 0.55M & 0.22/0.05 & 13.27/2.79/\textbf{0.90} \\
         4 & \xmark & -- & \cmark & 0.55M & 0.22/0.05 & 12.78/2.58/0.88 \\
         4 & \xmark & 1000 & \xmark & 0.45M & 0.22/0.00 & 11.24/2.42/0.86 \\
         \bottomrule
    \end{tabular}
    \label{tab:ablation}
\end{table}

We also conduct an ablation study to analyze the effect of the following design choices in DFSNet: feature partitioning; no PS in group GRU; normalization with sLN; and inclusion of CI through the average-group-concatenate scheme. Table \ref{tab:ablation} reports the ablation results on the entire test set. We note that feature partitioning without PS is highly effective in reducing model size and overall GMAC/s without negative impact on performance. We also verify that the use of sLN improves results by a noticeable margin, with $R = 1000$ attaining the best trade-off between performance and memory cost. Finally, we confirm that despite its low memory and computational cost, CI is indeed effective.\par

\section{Conclusion}
This paper proposed DFSNet, a steerable neural beamformer invariant to microphone array configuration for real-time SE. In contrast to conventional FS methods, DFSNet performs a channel alignment procedure prior to applying the FS operation, which simplifies the beamforming task into the estimation of DS clean reverberant speech. The proposed model incurs low latency, distortion, and memory and computational burden, making it suitable for hearing aid applications. Comparison with SOTA revealed that DFSNet outperforms causal methods in perception and intelligibility by a large margin. Additionally, we noted that DFSNet outperforms MVDR and approaches the performance of the noncausal FaSNet-TAC.\par

\section{Acknowledgements}

\ifdfsnetfinal
     This work was supported by the National Institute on Deafness and Other Communication Disorders (NIDCD) of the National Institutes of Health (NIH) under Award 5R01DC015430-05. The content is solely the responsibility of the authors and does not necessarily represent the official views of the NIH.\par
\else
     The authors would like to thank the funding organization of this research project.\par
\fi

\bibliographystyle{IEEEtran}
\bibliography{mybib}

\end{document}